# Field re-entrant superconductivity in Eu-doped infinite-layer nickelates

Mingwei Yang[1#], Jiayin Tang[1#], Xianfeng Wu[2#], Heng Wang[2,3*], Wenjing Xu[1,4], Haoliang Huang[2,3], Zhicheng Pei[1], Wenjie Meng[5], Guangli Kuang[5], Ming Yang[6], Jinfeng Xu[7], Sixia Hu[2], Junfeng Wang[6], Liang Li[6], Ze Wang[5], Chuanying Xi[5], Li Pi[5,7], Qingyou Lu[5,8,9], Ziqiang Wang[10], Qikun Xue[2,3,11], Zhuoyu Chen[2,3*], Danfeng Li[1,4*]

[1]*Department of Physics, City University of Hong Kong, Kowloon, Hong Kong SAR, China.*
[2]*State Key Laboratory of Quantum Functional Materials, Department of Physics, and Guangdong Basic Research Center of Excellence for Quantum Science, Southern University of Science and Technology, Shenzhen 518055, China.*
[3]*Quantum Science Center of Guangdong-Hong Kong-Macao Greater Bay Area, Shenzhen 518045, China.*
[4]*Shenzhen Research Institute of City University of Hong Kong, Shenzhen 518057, China.*
[5]*Anhui Key Laboratory of Low-Energy Quantum Materials and Devices, High Magnetic Field Laboratory, HFIPS, Chinese Academy of Sciences, Hefei 230031, China.*
[6]*Wuhan National High Magnetic Field Center and School of Physics, Huazhong University of Science and Technology, Wuhan 430074, China*
[7]*Institute of High Energy Physics, Chinese Academy of Sciences, Beijing 100049, China.*
[8]*Hefei National Research Center for Physics Sciences at the Microscale, University of Science and Technology of China; Hefei 230026, China.*
[9]*Hefei National Laboratory, University of Science and Technology of China, Hefei 230026, China.*
[10]*Department of Physics, Boston College, Chestnut Hill, MA, USA.*
[11]*Department of Physics, Tsinghua University, Beijing 100084, China.*
[#]*These authors contributed equally*: Mingwei Yang, Jiayin Tang, Xianfeng Wu.

[*]*Corresponding author*: wangheng@quantumsc.cn, chenzhuoyu@sustech.edu.cn, danfeng.li@cityu.edu.hk.

**Summary**

Intertwined superconducting and magnetic orders may give rise to exotic quantum phases[1–7], including field-induced and re-entrant superconductivity[8–10]. However, such magnetism-enhanced superconductivity has remained elusive in superconductors with higher transition temperatures[1–3]. While infinite-layer nickelates represent a new class of unconventional superconductors[11–20], the impact of rare-earth magnetism on superconducting properties remains largely unexplored. Here, we show that Eu-doped infinite-layer nickelate $Sm_{0.95-x}Ca_{0.05}Eu_xNiO_2$ exhibits a magnetic-field-induced re-entrant superconducting phase in the Eu-rich over-doped regime. Zero-resistance transport and high-field diamagnetic screening confirm the superconducting nature of this phase, which emerges after the initial suppression of low-field superconductivity and remains robust across a broad range of temperatures, fields and field orientations. In the same doping range, we observe nonlinear Hall transport and hysteretic magnetoresistance, indicating the unconventional nature of the re-entrant behaviour. While partially consistent with a compensation mechanism between the Eu-derived exchange field and the applied field, our data reveal pronounced deviations from this model at the highest-doping levels. Our findings establish infinite-layer nickelates as a fertile platform for exploring magnetically driven high-field superconductivity in strongly correlated oxides.

## Main text

Magnetism and superconductivity, through their intricate interplay, serve as foundations for various unconventional quantum phases, most notably high-temperature superconductivity[1–3], chiral superconductivity[4,5] and topological superconductivity[6,7]. While magnetism typically destroys Cooper pairs through orbital and paramagnetic effects, rare circumstances allow their coexistence and even mutual enhancement[8,9]. Experimental evidence for such delicate balance has been prominently observed in heavy-fermion compounds, where magnetic fluctuations may mediate unconventional pairing and lead to phenomena such as field-induced superconductivity[10–23]. Paradigmatic examples include uranium-based superconductors such as $UTe_2$[15–18], $UGe_2$[19,20], and URhGe[21–23], in which re-entrant superconductivity and/or spin-triplet superconductivity may emerge in the presence of strong magnetic correlations[15–23].

While cuprate superconductors represent a paradigmatic unconventional superconducting system with intertwined superconducting and antiferromagnetic orders[1], magnetic-field-stabilised superconducting phases associated with ferromagnetic instabilities have remained elusive. This has left the universality of such exotic superconducting phases an open question. The infinite-layer nickelates have emerged as a promising platform that potentially bridges high-temperature superconductivity and magnetic instabilities[24–33]. These materials, with transition temperatures reaching up to ~ 35 K[32,33], resemble cuprates in electronic structure while retaining important differences, particularly in the role of rare-earth elements[34–38]. Recent progress has opened new possibilities[31–33]: $Eu^{2+}$ ions carry large magnetic moments (S = 7/2) that may introduce substantial magnetic interactions. How Eu substitutions modulate the interaction between local magnetic moments and itinerant superconductivity remains largely unexplored and may reveal novel quantum phases.

In this work, we systematically investigate the phase diagram of ~ 22-u.c.-thick $Sm_{0.95-x}Ca_{0.05}Eu_xNiO_2$ ($SCE_x$) thin films grown on $(LaAlO_3)_{0.3}(Sr_2TaAlO_6)_{0.7}$ (LSAT) (00l) substrates, across a wide range of Eu concentrations ($x \sim 0 – 0.60$; and up to 0.80 in non-superconducting samples), revealing the emergence of strong magnetic correlations that coexists with superconductivity in the heavily doped regime. Most remarkably, we discover magnetic-field-induced re-entrant superconductivity persisting to extremely high magnetic fields. We demonstrate that this exotic behaviour arises from the interplay between magnetic $Eu^{2+}$ and superconducting Ni-O planes. While partially captured by the Jaccarino-Peter (J-P) compensation effect, our observations exhibit distinct departures suggesting the possibility of magnetically mediated superconducting pairing. These findings establish infinite-layer nickelates as a unique materials platform that brings together concepts from high-temperature and heavy-fermion superconductivity, opening entirely new avenues for engineering exotic quantum phases in strongly correlated materials.

Multiple sets of $SCE_x$ samples with varied Eu concentration were prepared (for sample growth, see Methods). X-ray diffraction data for representative precursor perovskite and product infinite-layer phases are shown in Extended Data Fig. 1, which indicates overall good crystallinity. The temperature-dependent resistivity, $\rho(T)$, normalised by its room-temperature value $\rho$(300 K), for one set of samples with varying $x$ is plotted in Fig. 1a (Individual $\rho$-$T$ curves are presented in Extended Data Fig. 2.). An interesting observation can be immediately made for $x = 0$ and $x < 0.10$: $SCE_x$ samples show an insulating behaviour with increasing $\rho(T)$ as temperature decreases. This is in direct contrast to the metallic behaviour (or superconducting[30,39–41]) as the precursor to a low-temperature weakly insulating state observed across other doped infinite-layer nickelate systems[25,26,42], despite a 0.05 Ca doping introduced even for $x = 0$ in $SCE_x$.

As $x$ increases, a metallic behaviour with a slight resistivity upturn at low temperatures accompanies the occurrence of the superconducting transition, first manifested as a downturn in resistivity ($x = 0.10$), before unveiling full transitions to a zero-resistance ground state with the superconducting transition temperature $T_c$ following a non-monotonic trend ($x = 0.10 \sim 0.45$). Upon further doping, $T_c$ vanishes, and the system becomes a canonical metal in almost the entire temperature range, with a clear change in the slope of the $\rho(T)$ curves. At these high doping levels ($x \geq 0.45$), a mild resistivity upturn is usually observed, indicative of disorder. The upturn temperature points ($T_{upturn}$) of the low-temperature resistivity are correspondingly plotted in Fig. 1c. The Eu-*M* edge spectrum (Fig. 1b) measured by X-ray absorption spectroscopy (XAS) exhibits features of mixed $Eu^{2+}/Eu^{3+}$ states, indicating that the oxidation state of Eu in infinite-layer nickelates is higher than +2—in agreement with prior reports for $Nd_{1-x}Eu_xNiO_2$[31]. In the inset of Fig. 1b, the Eu-$M_5$ edge peak evolution reveals a distinct low-energy shift with increasing $x$, consistent with an increase in the proportion of $Eu^{2+}$. Notably, this observation confirms that the $Eu^{2+}/Eu^{3+}$ ratio is not fixed across samples of varying Eu doping levels: it rises with increasing Eu doping concentration[43].

A colour map is shown in Fig. 1c, made of a contour plot of the $x$-dependent normalised $\rho(T)$ data ($x = 0.10 \sim 0.60$) for $SCE_x$, showcasing a robust 'dome'-shape superconducting regime. The maximum $T_c$ (onset) is ~ 32 K. The doping range in which superconductivity is present spans $0.08 < x < 0.45$, broader than that reported for the $Nd_{1-x}Eu_xNiO_2$ system ($0.15 < x < 0.40$)[31]. In addition to the expanded superconducting phase diagram, we observed that the samples retain excellent phase purity even under heavy doping levels of $x = 0.60$, with crystallinity maintained across the entire doping regime. As demonstrated in Supplementary Fig. S1, the $x = 0.80$ sample exhibits a good quality, although it lies much beyond the superconducting dome and is not the focus of this study.

The role of Eu dopants can be further examined particularly on the 'heavy-doping' side of the phase diagram and their possible contribution to the (magneto-)transport properties. We first performed Hall effect measurements on a series of $SCE_x$ samples. The measured $\rho_{yx}(H)$ data at different $T$ for a set of representative samples with different $x$ are shown in Extended Data Fig. 3. $R_H(T)$ was simply extracted from linear fits to the Hall resistivity of the entire field range, $\rho_{yx}(H)$, at various temperatures. For $0.28 \leq x \leq 0.60$, a sign change in $R_H(T)$ is observed. These features are consistent with the doping dependence reported in the literature[25,26,29,42]. For the 'heavily-doped' samples ($x > 0.32$), pronounced nonlinear characteristics emerge in $\rho_{yx}(H)$ at low temperatures after subtraction of the linear component, as exemplified by $x = 0.34$ in Fig. 2a and Extended Data Fig. 4. This nonlinear transport behaviour exhibits features distinct from those observed in multiband iron-based superconductors and two-dimensional electron gas systems[44,45], characterized by the emergence of a saturation plateau in $\rho_{yx}(H)$ under high magnetic fields (the temperature dependence of the saturation resistance value is shown in the insets of Fig. 2a, as an example). Moreover, a magnetic hysteresis in low-temperature magnetoresistance (MR) is also observed (Extended Data Fig. 5), as exemplified by the $x$ = 0.34 sample in Fig. 2b. The inset of Fig. 2b displays data from positive and negative field sweeps, and the inset in the top-left corner shows the temperature dependence of the peak value $R_{hys}$ of the hysteresis curve, consistent with the onset of additional magnetic correlations or time-reversal-symmetry-breaking in this regime. Notably, the presence of a large amount of $Eu^{2+}$ dopants in these compositions may induce magnetic interactions and significant spin-orbit coupling (SOC) that potentially contribute to the observed nonlinear $\rho_{yx}(H)$ and hysteresis in MR.

The response to magnetic field of the 'over-doped' $SCE_x$ was further tested on $SCE_{0.34}$, with a thickness of ~ 54 u.c. As shown in Fig. 2c, the MR exhibits a non-monotonic correspondence while $\mu_0H$ changes from 0 to 9 T, illustrating a clear field re-entrant behaviour. At $T$ = 2 K, a zero-resistance state was observed under $\mu_0H \geq$ ~ 6.23 T. More datasets of magnetic-field response for the 'over-doped' $SCE_x$ are presented in Extended Data Fig. 6. To elucidate the nature of the zero-resistance state and confirm the high-field re-entrant superconductivity, two-coil mutual inductance experiments were performed on this very sample. A clear diamagnetic response under 12 T can be observed (Fig. 2d), which, together with the signal at zero field, confirms the existence of two separate superconducting regimes. This superconducting-normal-superconducting transition largely resembles the re-entrant superconducting phase(s) that were discovered in heavy-fermion superconductors[15–18,20–23], Chevrel phase compounds[10,46] and organic conductors[11,14,47–50], which were ascribed to the interplay of (local) magnetism and (itinerant) superconductivity. Notably, evidence of spin-triplet superconducting pairing has been experimentally identified in heavy-fermion superconductors[15,21,51].

Attempting to fully unveil the behaviour of the re-entrant superconducting state, additional MR measurements were conducted on $SCE_{0.34}$ sample under the ultrahigh magnetic fields provided by a high-field magnet facility (see Methods for details). The magnetoresistance behaviour of the $SCE_{0.34}$ sample under different temperatures with magnetic fields applied out of the sample plane ($H//c$) and with varying angles (the polar angle $\theta$, angle between the external field and the $c$ axis) is presented in Figs. 2e and 2f. The $T$-dependent measurements reveal that, at 1.9 K, the second superconducting phase maintains for a wide magnetic field range of ~ 15 – 30 T. Notably, this field-induced re-entrant superconductivity persists up to 10 K, indicating the exceptional robustness of the phenomenon. As explicitly demonstrated in Fig. 2f, field-induced re-entrant superconductivity persists across an exceptionally broad angular range of up to 70° – substantially exceeding the narrow angular tolerances reported for Chevrel phase compounds and organic conductors (~ 2 – 10°)[11,14,47], and heavy-fermion superconductors (~ 30°)[15,16,22,23], highlighting the unprecedented angular robustness of the re-entrant state in this material system.

To systematically investigate the magnetic-field-enhanced superconductivity, we conducted comprehensive ultrahigh-field measurements on a series of over-doped samples ($x$ = 0.30 – 0.40). Complete datasets illustrating consistent behaviour are presented in Extended Data Fig. 7. All $SCE_x$ ($x$ = 0.32, 0.33, 0.36, 0.38, 0.40) compounds exhibit high-field-induced re-entrant superconductivity, and this re-entrant behaviour is observed across a wide angular range from 0° to 70°, even persisting at 90° (Extended Data Figs. 8 and 9). As an example, Extended Data

Fig. 7b displays the measured MR of a $SCE_{0.32}$ at different temperatures under a perpendicular field up to 35 T. Similar to our observations for $x = 0.34$, a field-induced superconducting state is fully developed at sufficiently high magnetic fields: the initial field applied mainly suppresses superconductivity with $\rho(H)$ increasing from zero to finite values; above a critical field strength (3.23 T), $\rho(H)$ drops again towards zero-resistance state and reaches a true zero around 13.04 T at $T = 3$ K. This high-field zero-resistance state sustains above ~ 32 T (at $T = 1.7$ K). This re-entrant behaviour – also observed for $\theta \leq 70°$ – exhibits exceptional field robustness. Notably, this field-enhanced superconductivity phenomenon was not observed at lower doping levels (for instance, $x = 0.30$; see Extended Data Fig. 7a), which suggests that Eu doping serves as a critical parameter for this high-field enhancement.

Figures 3a-e systematically map the superconducting phase evolution of $SCE_x$ with $x = 0.30$, 0.33, 0.36, 0.38 and 0.40 by constructing *H-T-R* phase diagrams for *H*//*c*. For $SCE_{0.30}$, superconductivity is monotonically suppressed by field without resolved re-entrant features, consistent with Extended Data Fig. 7a. When $x$ increases to 0.33, 0.36, and 0.38, a second high-field superconducting region emerges: $T_c$ is suppressed at the intermediate field strength and then recovers at higher fields, yielding a clear superconducting-normal-superconducting sequence in the *H-T-R* plane. This places the onset of the re-entrant regime on the over-doped side near $x \sim 0.32$ in the present dataset. With increasing $x$, the high field 'dip' deepens and broadens; by $x = 0.38$, the transition splits into two well-separated superconducting regions. Note that although the data of $T_{c0}$ at high fields is lacking for $SCE_{0.40}$, the re-entrant feature can still be clearly identified, as shown in Extended Data Fig. 7f.

A plausible origin of this behaviour is the J-P compensation effect[8]: the internal exchange field ($H_J$) generated by the $Eu^{2+}$ moments partially cancels the externally applied field ($H_{ext}$). As sketched in Supplementary Fig. S2, the effective field acting on Cooper pairs is $H_{tot} = H_J + H_{ext}$; re-entrance arises when $|H_{tot}|$ falls below the intrinsic upper critical field, $H_{c2}$. To examine this phenomenology, we model $H_{c2}(T)$ using Fisher's extension[52] of the Werthamer-Helfand-Hohenberg (WHH) theory[53], including orbital and Pauli pair-breaking, spin-orbit scattering and an exchange field from Eu moments. The dashed curves in in Figs. 3a-e show fits obtained using this unmodified baseline form. This approach captures the overall evolution from monotonic suppression at $x = 0.30$ to clear re-entrant behaviour at intermediate Eu concentrations; but systematic deviations remain, especially in the low-field region for $x = 0.33$ – 0.38 and more strongly for $x = 0.40$. For completeness, we also compare these baseline fits with an empirical variant in which a scaling factor of 2 is introduced in the argument of the Brillouin function, as shown in Extended Data Fig. 8 and discussed in Methods. The baseline fits in Figs. 3a-e yield an exchange field $H_J$ of order - 40 T for the re-entrant compositions, together with a moderate spin-orbit scattering parameter $\lambda_{so}$ that increases with $x$, and a Maki parameter $\alpha$ of order 3 – 5 (Fig. 3f; detailed values are summarized in Extended Data Table 1). Although these values should be regarded as phenomenological fit parameters, their systematic evolution supports the relevance of exchange-field compensation in the intermediate-doping regime. In particular, the similar $\alpha$ values obtained for $x = 0.30$ and $x = 0.38$, together with the much larger $|H_J|$ in the re-entrant samples, suggesting that $|H_J|$ is the parameter most closely correlated with onset of re-entrance. When $x = 0.36$ – 0.38, $H_J$ becomes comparable in scale to $H_{c2}^{\mathrm{orb}}(0)$, which maximizes the compensation window and coincides with the strongest re-entrant features. By contrast, the clear mismatch between the baseline fit and the data for $x = 0.40$ indicates that additional physics is likely involved in that composition.

The angular dependence provides further information regarding the mechanism of the re-entrant behaviour. Figures 3g-j illustrate the evolution of the re-entrant feature in $SCE_{0.33}$, $SCE_{0.36}$, $SCE_{0.38}$, and $SCE_{0.40}$ as a function of $\theta$. Qualitatively, increasing Eu doping systematically broadens the angular window for field-induced re-entrant superconductivity from ~ 70° in $SCE_{0.33}$ to nearly 90° in $SCE_{0.38}$ and $SCE_{0.40}$. Concomitantly, the critical magnetic field required for the low-field superconducting state to collapse to a normal state gradually shifts toward zero field. For further analysis, using $SCE_{0.36}$ as a representative case, we reconstruct $H_{c2}(\theta)$ within the same WHH-J-P framework, using the empirical modified Brillouin term introduced in Methods, and allowing $H_J$, $\lambda_{so}$ and $\alpha$ to vary weakly with $\theta$ (see Methods). The resulting multi-branched $H_{c2}(\theta)$ cannot be captured by either Tinkham or anisotropic Ginzburg-Landau expressions. Instead, it is qualitatively described by a microscopic picture, in which an almost isotropic Eu-derived exchange field combines with angle-dependent orbital and Pauli pair-breaking, tuned by spin-orbit scattering (discussed in Supplementary Fig. S3). This isotropic assumption is consistent with the system tending toward a three-dimensional superconducting state as Eu doping increases[33]. Notably, as shown in Figs. 4a-c and Extended Data Figs. 7 and 8, for the $H\perp c$ configuration, the Eu doping level of $x$ ~ 0.36 defines the onset threshold of the re-entrant phase in the over-doped region.

Although the J-P compensation provides a phenomenological description of the re-entrant superconductivity in $SCE_{0.33}$, $SCE_{0.36}$ and $SCE_{0.38}$, $SCE_{0.40}$ shows clear deviations from this framework. As illustrated in Figs. 3j and 4c, superconductivity in $SCE_{0.40}$ is initially suppressed by applied field but subsequently re-established and even enhanced—resembling the field-induced re-entrance discussed above. However, the high-field superconducting phase is markedly more robust than the initial low-field state, and superconductivity collapses at substantially lower fields for $H//c$ (< 1 T; Fig. 3e and Extended Data Fig.7f), indicating a strong role for orbital pair breaking. These features are not captured by the simple J-P compensation picture alone and point to additional physics in the highest-doping regime. This behaviour invites comparison with other field-enhanced superconducting systems in which unconventional pairing has been discussed, but establishing the pairing symmetry here will require further experiments.

As evidenced by the angular-dependent measurements for $x = 0.40$ in Fig. 3j, the collapse of bifurcated angular windows – unlike those observed at $x = 0.38$ – signals the possible breakdown of J-P compensation effects. Furthermore, the field-induced re-entrant superconductivity persists across the full angular range (0° – 90°), a phenomenon also observed in $x = 0.38$ samples, as shown in Extended Data Fig. 9c. This consistent trend suggests a continuous evolution of the system with increasing Eu doping level, raising a critical question regarding whether orbital de-pairing effects should also be taken into consideration in the smaller-$x$ regime.

The main observations and key ingredients relevant to superconductivity and the re-entrant phenomenon in this intriguing IL nickelate system are then summarised into a rich phase diagram, illustrated in Fig. 4d, with key temperature scales labelled. The diagram reveals a systematic evolution from insulating behaviour in the underdoped regime to metallic superconductivity, culminating in a low-temperature upturn in resistivity in the over-doped region. Critically, the re-entrant phenomenon persists over a broad over-doped range ($x \sim 0.32$ to 0.40), demonstrating its robustness in the IL nickelate system. The phase diagram further incorporates characteristic temperatures where the suppression of nonlinear Hall saturation resistance, combined with the observed magnetoresistance hysteresis, are consistent with additional magnetic correlations and possible time-reversal-symmetry breaking, suggesting emergent long-range magnetic order or strong spin-orbit coupling effects that warrant future investigation.

The lattice compatibility of Eu in the A site of $SCE_x$ largely enabled a unique materials platform for the investigation of a superconducting phase across a broad range of doping concentrations. This assists in overcoming the solubility limit encountered for heavily doped materials (e.g. Sr and Ca as dopants). The introduction of Eu ions enables the observation of a high-field superconducting state, promoted by the magnetic interactions from an intriguing magnetic phase on the over-doped side of the phase diagram.

It is worth noting some interesting and unique aspects of this re-entrant superconducting behaviour. First, this field-induced state occurs with a $T_c$ that is significantly higher than that of previously reported re-entrant systems[15–23]. Next, despite a generally large $H_{c2}$ of the IL nickelate system, the full superconducting state induced at high magnetic fields can be observed at a field as low as ~ 6 T, making it easily accessible for investigation. In addition, distinct from the usually observed re-entrant superconducting state under in-plane parallel field in other layer-structured material systems[11,14,47], in $SCE_x$, field-induced re-entrant superconductivity was observed at its strongest under perpendicular magnetic field with respect to the electron transport plane. A systematic angle-dependent re-entrant behaviour can be clearly seen with $\theta$ in a remarkably wide range of ~ 0 – 90º (Fig. 3, Fig. 4c), larger than the ~ 2° – 30° windows in other systems[11,14–16,22,23,47,54]. This is also in stark contrast to the re-entrant superconducting state in two-dimensional superconductors, where the orbital de-pairing is quenched or suppressed and the Zeeman effect induced by in-plane magnetic field is largely compensated[8,14,47]. In particular, for the $SCE_{0.40}$ sample under perpendicular magnetic field, orbital effects cannot be neglected—beyond the J-P scenario. Last, the characteristic field scales for the re-entrant superconducting phase are in line with the emergence of pronounced hysteretic magnetoresistance and a sizable non-linear Hall effect (Figs. 2a and 2b; Extended Data Figs. 3-5,). This indicates that the possible magnetic interactions may play a central role in stabilizing the re-entrant state. In this regard, the enhancement of the superconducting state at high fields presents an underlying mechanism of magnetically mediated superconductivity, analogous to that in heavy fermion superconductor. This evokes the tempting possibility of achieving a spin-triplet state, which calls for further exploration. We note, however, that the present study does not determine the presence of long-range magnetic order or its detailed structure. To address these issues, additional spectroscopic and local probes, such as muon spin relaxation, x-ray magnetic circular dichroism, resonance (in-)elastic x-ray scattering, are needed.

In summary, we establish the Eu-doping phase diagram of $Sm_{0.95-x}Ca_{0.05}Eu_xNiO_2$ thin films over a wide doping range, and identify a re-entrant superconducting state under ultrahigh magnetic fields in the over-doped regime. The accompanying nonlinear Hall response and hysteretic magnetoresistance in the same regime suggest the presence of unconventional magnetic and electronic coupling. While a field-compensation phenomenology qualitatively captures the intermediate-doping behaviour, the observation at highest-doping levels points to additional ingredients beyond this simple picture. These collective findings position infinite-layer nickelates as a versatile platform for investigating how rare-earth magnetism can reshape superconductivity in a correlated oxide with a high transition temperature.

**Main figure legends**

**Fig. 1 | Eu doping dependent resistivity, electronic state and phase diagram of $Sm_{0.95-x}Ca_{0.05}Eu_xNiO_2$ ($SCE_x$).** **a**, The normalised resistivity ($\rho(T)/\rho(300\text{ K})$) as a function of temperature for a set of representative samples with different $x$ ($0 \leq x \leq 0.60$). **b**, X-ray absorption spectroscopy of $SCE_x$ ($x$ = 0, 0.10, 0.22, 0.32, 0.40, 0.50, 0.60) showing the $M$ edge of Eu. Each curve was normalised with reference to its respective Eu-$M_5$ peak intensity. Inset: The peak positions of the Eu-$M_5$ edge extracted from the main panel of **b** plotted against $x$. A

clear decrease in the energy values can be observed as $x$ increases, showcasing a doping effect. **c**, A contour plot with the colour code representing the normalised resistivity ($\rho(T)/\rho(150\ \mathrm{K})$) values as a function of $x$, showing the superconducting phase. $T_{c,\mathrm{onset}}$, the onset of superconductivity, defined as the temperature below which the $\rho$ data start to drop below a linear extrapolation of the normal-state $\rho(T)$; $T_{\mathrm{upturn}}$, the characteristic temperature below which the resistance shows a small upturn.

**Fig. 2 | Hall effect measurements and magnetotransport of over-doped SCE$_x$ samples illustrating the re-entrant superconducting state. a**, Nonlinear Hall resistance $\Delta R_{xy}$ of an SCE$_{0.34}$ sample obtained after subtracting the linear component. Inset, temperature dependence of the saturation resistance; grey curves are guides to the eye. **b**, Magnetoresistance hysteresis of an SCE$_{0.34}$ sample at different temperatures. Right inset, data measured at 1.9 K under positive and negative field sweeps, where red and black denote field-up and field-down sweeps, respectively. Left inset, temperature dependence of the peak value $R_{\mathrm{hys}}$. **c**, $\rho(H)$ of an SCE$_{0.34}$ sample (~ 54 u.c.). At 2 K, $\rho$ returns to zero above 6.23 T. Inset, enlarged view of the 2 K data. **d**, Mutual-inductance measurements under various magnetic fields ($H//c$). The real ($\mathrm{Re}(V\mathrm{p})$) and imaginary ($\mathrm{Im}(V\mathrm{p})$) signals show a diamagnetic response at $\mu_0 H$ = 0 T, which is absent at $\mu_0 H$ = 3 T. At $\mu_0 H$ = 9 T and 12 T, the drop in $\mathrm{Re}(V\mathrm{p})$ and the peak in $\mathrm{Im}(V\mathrm{p})$ indicate diamagnetic screening associated with re-entrant superconductivity. **e**, $\rho(H)$ up to 60 T for $H//c$ at different temperatures for SCE$_{0.34}$. **f**, $\rho(H)$ up to 60 T at different $\theta$ angles for SCE$_{0.34}$ measured at 4.2 K. $\theta$ is the angle between $H$ and the sample surface normal. The polar angle $\theta$ is determined from the perpendicular magnetic-field component captured by a secondary pick-up coil embedded in the rotor, yielding the field-to-sample-normal angle.

**Fig. 3 | Magnetotransport of 'over-doped' SCE$_x$ samples illustrating the re-entrant superconducting state. a-e**, Contour maps of superconductivity of SCE$_x$ with $x$ = 0.30, 0.33, 0.36, 0.38 and 0.40, constructed from $H$-$T$-$R$ phase diagrams with $H//c$. The colour represents the resistance normalized by the high-field normal-state resistance. Open circles denote the $0.5R_n$ criterion. Dashed lines show fits using the unmodified Fisher's extension of the Werthamer-Helfand-Hohenberg (WHH) theory. A comparison with an empirical modified Brillouin form is provided in Extended Data Fig. 8. **f**, Doping dependence of the phenomenological fit parameters, including the spin-orbit scattering parameter $\lambda_{so}$, Maki parameter $\alpha$, and exchange field $H_J$, obtained from the baseline fits in **a-e**. The pink shaded regions denote the re-entrant superconducting region. **g-j**, Contour maps of superconductivity for SCE$_x$ with $x$ = 0.33, 0.36, 0.38 and 0.40, constructed from $H$-$\theta$-$R$ phase diagrams measured at $T$ = 4.2 K for $x$ = 0.33 and 0.36, $T$ = 1.5 K for $x$ = 0.38, and $T$ = 0.3 K for $x$ = 0.40. The colour represents the resistance normalized by the high-field normal-state resistance. Open circles denote the $0.5R_n$ criterion.

**Fig. 4 | Magnetotransport of SCE$_x$ showing the re-entrant superconducting state and the phase diagram of $Sm_{0.95-x}Ca_{0.05}Eu_xNiO_2$. a-c**, Contour plot of $\rho(H)$ measured with $H\perp c$ for SCE$_x$ with $x$ = 0.36, 0.38 and 0.40, respectively, at different temperatures. Dashed lines show fits using the unmodified Fisher's extension of the Werthamer-Helfand-Hohenberg (WHH) theory. **d**, Superconducting phase diagram of $Sm_{0.95-x}Ca_{0.05}Eu_xNiO_2$. Here, $T_{c,\mathrm{onset}}$ denotes the onset temperature of the superconducting transition; $T_{\mathrm{upturn}}$ is the characteristic temperature

below which resistance exhibits a weak upturn; $T_{nonlinear}$ represents the temperature below which the Hall resistance displays nonlinear behaviour. Inset, lattice structure of infinite-layer $SCE_x$.

## Methods

### Thin-film growth and preparation

Polycrystalline $Sm_{0.95-x}Ca_{0.05}Eu_xNiO_y$ ceramic targets were synthesised by mixing stoichiometric ratios of $Sm_2O_3$, $Eu_2O_3$, $CaCO_3$, and NiO powders, followed by decarbonization at 1250 °C for 12 hours. After that, the targets were ground, re-pelletised and then sintered twice at 1300 °C and 1270 °C for 12 hours each. ~ 22-u.c.-thick $Sm_{0.95-x}Ca_{0.05}Eu_xNiO_3$ epitaxial films were deposited on 5 × 5 mm² LSAT (001) substrates via KrF excimer laser ($\lambda$ = 248 nm) enabled pulsed laser deposition (PLD). Growth was performed at 600 °C under an oxygen partial pressure of 100 mTorr, with a laser fluence of 2.6 – 2.8 J/cm² and a repetition rate of 2 – 3 Hz. ~ 5 u.c.-thick $SrTiO_3$ capping layer was subsequently grown using a fluence of 0.6 J/cm². Topotactic reduction was carried out *in situ* inside a vacuum chamber using $CaH_2$ powders: The infinite-layer $Sm_{0.95-x}Ca_{0.05}Eu_xNiO_2$ phase formation was tracked by real-time monitoring of the sample's minimum resistance through a two-probe configuration. The process was conducted at a thermocouple-measured temperature range of ~ 270 – 310 °C for 0.5 - 1.5 hours. A small $Ca^{2+}$ substitution on the rare-earth site was shown to markedly facilitate the topotactic reduction from the perovskite to the infinite-layer phase, improving structural integrity and superconducting properties.

### X-ray diffraction characterisation

The X-ray diffraction $\theta - 2\theta$ symmetric scans of the films were obtained by a Rigaku SmartLab (9 kW) high-resolution X-ray diffractometer with the wavelength of the X-ray being 0.154 nm. The XRR fits in Extended Data Fig. 1 were obtained using a standard Parratt formalism as implemented in the commercial analysis package (Rigaku GlobalFit). ~ 5 u.c.-thick $SrTiO_3$ capping layer was included in the XRR model as the topmost layer. Its thickness and roughness were refined together with those of the infinite-layer film.

### Lab-based transport measurements

Wire connections for electrical transport measurements using the standard four-probe method were made with aluminium ultrasonic wire-bonded contacts. Resistivity and Hall-effect measurements were performed at temperatures down to 2 K and under magnetic fields up to 14 (or 9) T using Physical Properties Measurement Systems from Quantum Design Inc. Two-coil mutual inductance experiments were conducted with the driving and pickup coils aligned vertically above and below the thin film samples.

### High-magnetic-field measurements

High-magnetic-field magnetoresistance measurements were performed at the Chinese High Magnetic Field Laboratory (CHMFL) in Hefei. The experiments utilized a 35 T water-cooled magnet and a 45.2 T hybrid magnet[55], which are equipped with a $^3He$ cryostat.

Magnetoresistance was measured using SR830 lock-in amplifiers with a built-in a.c. current source. Part of the high-field magnetoresistance measurements up to 60 T were performed using the standard four-wire technique with high-frequency (100 kHz) AC excitations at 390 μA, in a pulsed magnet with a duration of 160 ms at the Wuhan National High Magnetic Field Center (WHMFC).

**X-ray absorption spectroscopy measurements**

The soft X-ray absorption spectroscopy (XAS) measurements at the Eu-*M* edge for the prepared samples were conducted at room temperature using total electron yield (TEY) mode with the X-ray beam incident at 90° relative to the sample surface (i.e. 0° relative to the surface normal) on the BL08U1A soft X-ray spectro-microscopy beamline at the Shanghai Synchrotron Radiation Facility (SSRF).

**Theoretical modelling**

The temperature dependence of the upper critical field, $H_{c2}(T)$, presented in Figs. 3a-d, is fitted by the following implicit equation:

$$\ln\frac{1}{t} = \left(\frac{1}{2}+\frac{i(\lambda_{\mathrm{so}}-\lambda_m)}{4\gamma}\right)\psi\left(\frac{\frac{1}{2}+h+\lambda_m+\frac{1}{2}(\lambda_{\mathrm{so}}-\lambda_m)+i\lambda}{2t}\right)$$
$$+\left(\frac{1}{2}-\frac{i(\lambda_{\mathrm{so}}-\lambda_m)}{4\gamma}\right)\psi\left(\frac{\frac{1}{2}+h+\lambda_m+\frac{1}{2}(\lambda_{\mathrm{so}}-\lambda_m)-i\lambda}{2t}\right)-\psi\left(\frac{1}{2}\right),$$

where $\lambda = \sqrt{\alpha^2(h+h_J)^2-\frac{1}{4}(\lambda_{\mathrm{so}}-\lambda_m)^2}$ . $\lambda_{\mathrm{so}}$ and $\lambda_m$ are the spin-orbit and magnetic scattering parameter. The reduced parameters are defined as: $t = T/T_c$, $h = 0.281\, H_{c2}(T)/H_{c2}^{\mathrm{orb}}(0)$, $h_J = 0.281\, H_J(T)/H_{c2}^{\mathrm{orb}}(0)$, where $H_{c2}^{\mathrm{orb}}(0)$ is the orbital limited critical field at $T = 0$ K, Maki parameter $\alpha = \sqrt{2}\frac{H_{c2}^{orb}}{H_p}$ and $H_p$ is the Pauli-limited critical field. $\lambda_m$ is set to be 0 here. $\psi$ is the digamma function. $H_J(T)$ is described by a revised Brillouin function for the $Eu^{2+}$ ions. The Brillouin function is defined as:

$$\mu_0 H_J(x) = \frac{2J+1}{2J}\coth\left(\frac{2J+1}{2J}\,x\right)-\frac{1}{2J}\coth\left(\frac{x}{2J}\right),$$

where $J$ is the angular-momentum quantum number (7/2 for $Eu^{2+}$), $x = \frac{g_J\mu_B B}{k_B T}$ is the ratio of magnetic to thermal energy ($g_J$ is Landé g-factor, $\mu_B$ is Bohr magneton, $k_B$ is Boltzmann Constant). For comparison, we also consider an empirical modification in which an extra factor 2 is introduced in $x$. This modification causes $H_J$ to saturate more rapidly than in the standard Brillouin form and improves the description of the intermediate-doping data, but it is not universal and does not provide a satisfactory account for the $x = 0.40$ sample.

**Data and code availability statements**

The data that support the findings this study are available from doi.org/10.6084/m9.figshare.31861684.

**Acknowledgements**

We acknowledge the funding support from the National Natural Science Foundation of China (12174325), a Guangdong Basic and Applied Basic Research Grant (2023A1515011352) and a Shenzhen Science and Technology Program (JCYJ20250604191219025). The research was supported by research grants from the Research Grants Council (RGC) of the Hong Kong Special Administrative Region, China, under Early Career Scheme, General Research Fund and ANR-RGC Joint Research Scheme (CityU 21301221, CityU 11309622, CityU 11300923, CityU 11313325 and A-CityU102/23). Part of the work utilised the equipment support through a Collaborative Research Equipment Grant from RGC (C1018-22E). Part of the research work described in this paper was conducted in the JC STEM Lab of Energy and Materials Physics funded by The Hong Kong Jockey Club Charities Trust. Part of this work was supported by the National Key R&D Program of China (2024YFA1408101 and 2022YFA1403101), the Natural Science Foundation of China (92565303, 12504161, 92265112, 12374455 & 52388201), the Guangdong Major Project of Basic Research (No. 2025B0303000004), the Guangdong Provincial Quantum Science Strategic Initiative (GDZX2501001, GDZX2401004 and GDZX2201001), the Shenzhen Science and Technology Program (KQTD20240729102026004), and the Shenzhen Municipal Funding Co-Construction Program Project (SZZX2301004 and SZZX2401001). The high-magnetic-field work was supported by the National Key R&D Program of China (2023YFA1607701) and the National Natural Science Foundation of China (51627901). H.W. acknowledges the supported by the Interdisciplinary program of Wuhan National High Magnetic Field Center (Grant No.2025006), Huazhong University of Science and Technology (HUST) and thanks the staff members of the electrical transport station and the 65 T-magnet system at the Pulsed High Magnetic Field Facility in HUST, for providing technical support and assistance in data collection. Z.W. acknowledges the support of U.S. Department of Energy, Basic Energy Sciences Grant No. DE-FG02-99ER45747. We thank the staff members of the SMA System (https://cstr.cn/31125.02.SHMFF.SM2.SMA) at the Steady High Magnetic Field Facility, CAS (https://cstr.cn/31125.02.SHMFF), for providing technical support and assistance in data collection and analysis. We acknowledge the support from the Station of Quantum Materials.

**Author contributions**

M.W.Y., J.T., X.W. contributed equally to this work. M.W.Y., H.W., Z.C. and D.L. conceived the research project. M.W.Y., J.T. and W.X. grew the samples. W.X. and M.W.Y. synthesised polycrystalline targets. H.W., X.W., Z.P. and W.X. performed the in-situ reduction experiments.

H.W., X.W., W.X., M.W.Y., J.T., Z.P. and S.H. performed transport measurements. H.W. conducted the mutual inductance measurements. H.W., M.W.Y., and H.H. conducted the high field measurements in Hefei with the help of W.M., G.K., C.X., L.P., Z.W. and Q.L. H.W. and M.W.Y. conducted the high field measurements in Wuhan with the help of M.Y., J.W., and L.L. M.W.Y. and J.X. conducted the XAS measurements. Z.Q.W. provided theoretical interpretations. Q.X., Z.C., and D.L. acquired funding support. M.W.Y., J.T., H.W., Z.C., and D.L. wrote the manuscript with contributions from all authors.

**Author information**

Competing interests: The authors declare no competing interests.

Supplementary Information: The online version contains supplementary material available at (to be filled)

Correspondence and requests for materials should be addressed to Danfeng Li, Zhuoyu Chen or Heng Wang.

Peer review information.

**Extended data figure legends**

**Extended Data Fig. 1 | The structural characterisations of the samples. a** and **b**, X-ray diffraction (XRD) $\theta$–$2\theta$ symmetric patterns of $Sm_{0.73}Ca_{0.05}Eu_{0.22}NiO_{3(2)}$ and $Sm_{0.61}Ca_{0.05}Eu_{0.34}NiO_3$ thin films. The presence of prominent (00l) film peaks with finite-size oscillations and their peak positions all suggest good film crystallinity across wide ranges of doping and film thickness. c, X-ray reflectivity (XRR) patterns for $Sm_{0.61}Ca_{0.05}Eu_{0.34}NiO_3$ thin films with thicknesses of ~ 22 u.c. and ~ 54 u.c. In experiments, the infinite-layer phase samples with varying doping levels all have a thickness of ~ 22 u.c. The sample $SCE_{0.34}$ used for the mutual inductance two-coil measurements has a thickness of ~ 54 u.c. The XRR fits in Extended Data Fig. 1 were obtained using a standard Parratt formalism as implemented in the commercial analysis package (Rigaku GlobalFit). The ~ 5 u.c. $SrTiO_3$ capping layer was included in the XRR model as the topmost layer. Its thickness and roughness were refined together with those of the infinite-layer film.

**Extended Data Fig. 2 | Individual $\rho(T)$ curves of $Sm_{0.95-x}Ca_{0.05}Eu_xNiO_2$ ($SCE_x$) across all doping levels.** The inset of Panel **v** is the zoom-in curve at low temperatures, showing a small downturn in resistivity near $T = 2.3$ K.

**Extended Data Fig. 3 | Hall resistivity of $SCE_x$ samples across different regimes.** No sign change in Hall coefficient ($R_H$, slope of $\rho_{yx}(H)$) for $x$ lower than 0.24; Sign change in $R_H$ for $x$ over 0.24; Nonlinear Hall resistivity at low temperature for $x$ over 0.34.

**Extended Data Fig. 4 | Non-linear Hall resistance of SCE$_x$ samples, $x$ = 0.40, 0.50, 0.55 and 0.60.** The insets show the temperature dependence of the saturation resistance, with the grey curves serving as guides to the eye.

**Extended Data Fig. 5 | Magnetoresistance hysteresis of a SCE$_{0.38}$ sample at different temperatures.** The insets show the temperature dependence of the peak value $R_{hys}$ of the hysteresis curve, with the grey curves serving as a guide to the eye.

**Extended Data Fig. 6 | Magnetic-field responses of SCE$_x$ samples across different regimes.** Clear re-entrant behaviour starts to show up as $x$ exceeds 0.32.

**Extended Data Fig. 7 | High field $\rho(H)$ at different temperatures for SCE$_x$ samples under out-of-plane and in-plane fields.** Here, the $\rho(H)$ curves for SCE$_{0.30}$ show no re-entrant behaviour up to 60 T at $H//c$, clearly demarcating a lower limit of the re-entrant regime. Typical re-entrant behaviour is clearly exhibited for the composition range of $0.32 \leq x \leq 0.38$ at $H//c$. The $\rho(H)$ curves for SCE$_{0.36}$ show no re-entrant behaviour up to 60 T at $H\perp c$. The $\rho(H)$ curves of SCE$_{0.38}$ and SCE$_{0.40}$ under in-plane field show a sign of field enhanced superconductivity.

**Extended Data Fig. 8 | Empirically modified fits to the high-field superconducting phase diagrams. a-e**, *H-T-R* contour maps for SCE$_x$ with $x$ = 0.30, 0.33, 0.36, 0.38 and 0.40 for $H//c$, reproduced from the main analysis for direct comparison. **f**, Composition dependence of the corresponding fit parameters. The empirical modification improves the agreement for intermediate compositions ($x$ = 0.33 - 0.38). **g,h**, *H-T-R* contour maps for SCE$_x$ with $x$ = 0.36 and 0.38 for $H\perp c$. All black dashed lines show fits obtained after introducing an empirical factor of 2 in the argument of the Brillouin function.

**Extended Data Fig. 9 | $\rho(H)$ up to 45 T with different $\theta$ angles for SCE$_x$ samples.** $\theta$ is the angle between $H$ and the sample surface normal.

**Extended Data Table 1 | $H_{c2}^{orb}$ and phenomenological fit parameters for temperature-dependent $H_{c2}$ of different samples.** $H_J$, $\lambda_{so}$, and $\alpha$ are obtained from fits to the unmodified Fisher-extension WHH model. Values in parentheses were obtained using the empirical modified Brillouin form with an extra factor of 2 in the argument.

# Figures

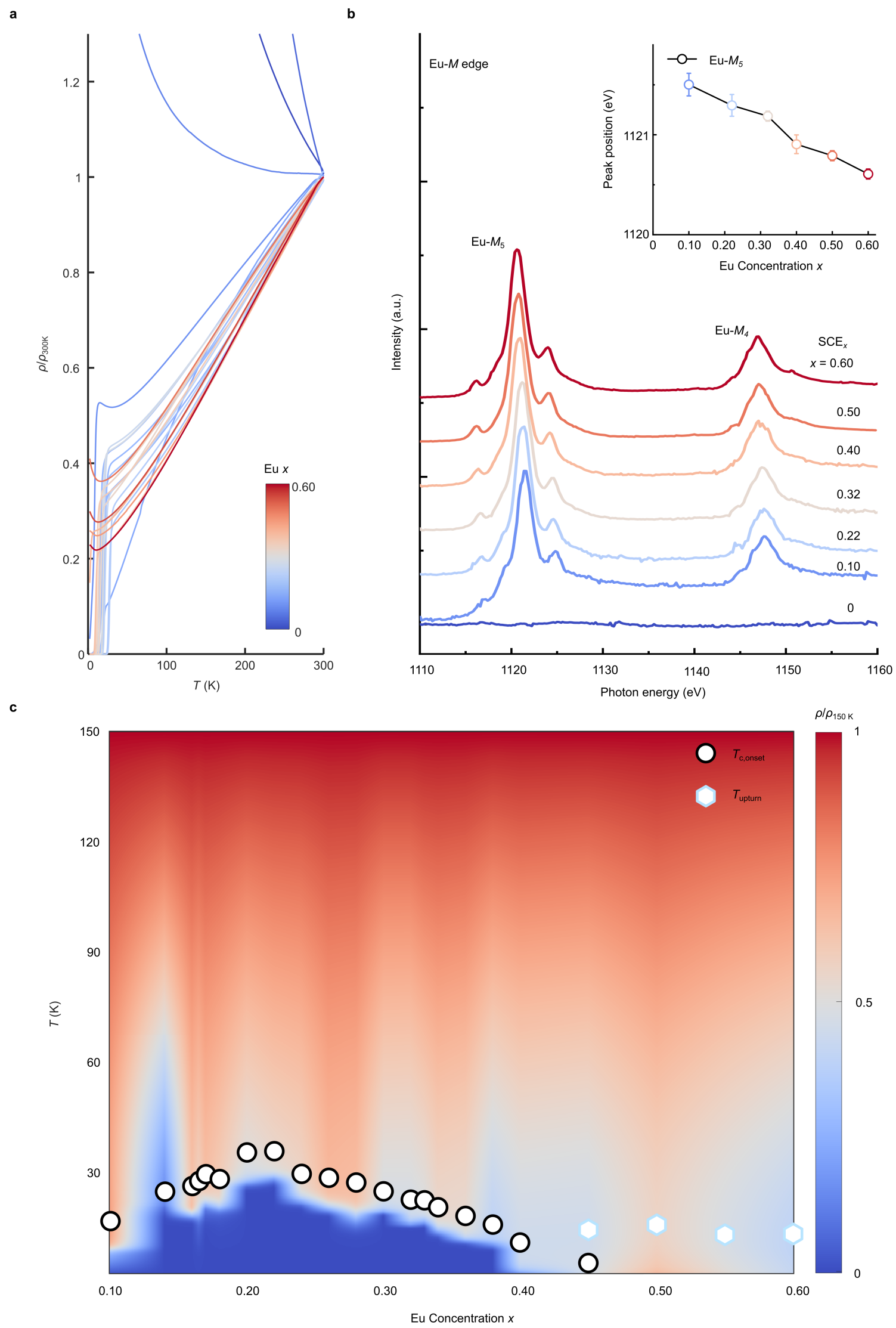

a
b
c
ρ/ρ300K
T (K)
Eu x
0.60
0
Eu-M edge
Eu-M5
Eu-M4
Intensity (a.u.)
Photon energy (eV)
SCEx
x = 0.60
0.50
0.40
0.32
0.22
0.10
0
Peak position (eV)
Eu Concentration x
T (K)
ρ/ρ150 K
Tc,onset
Tupturn

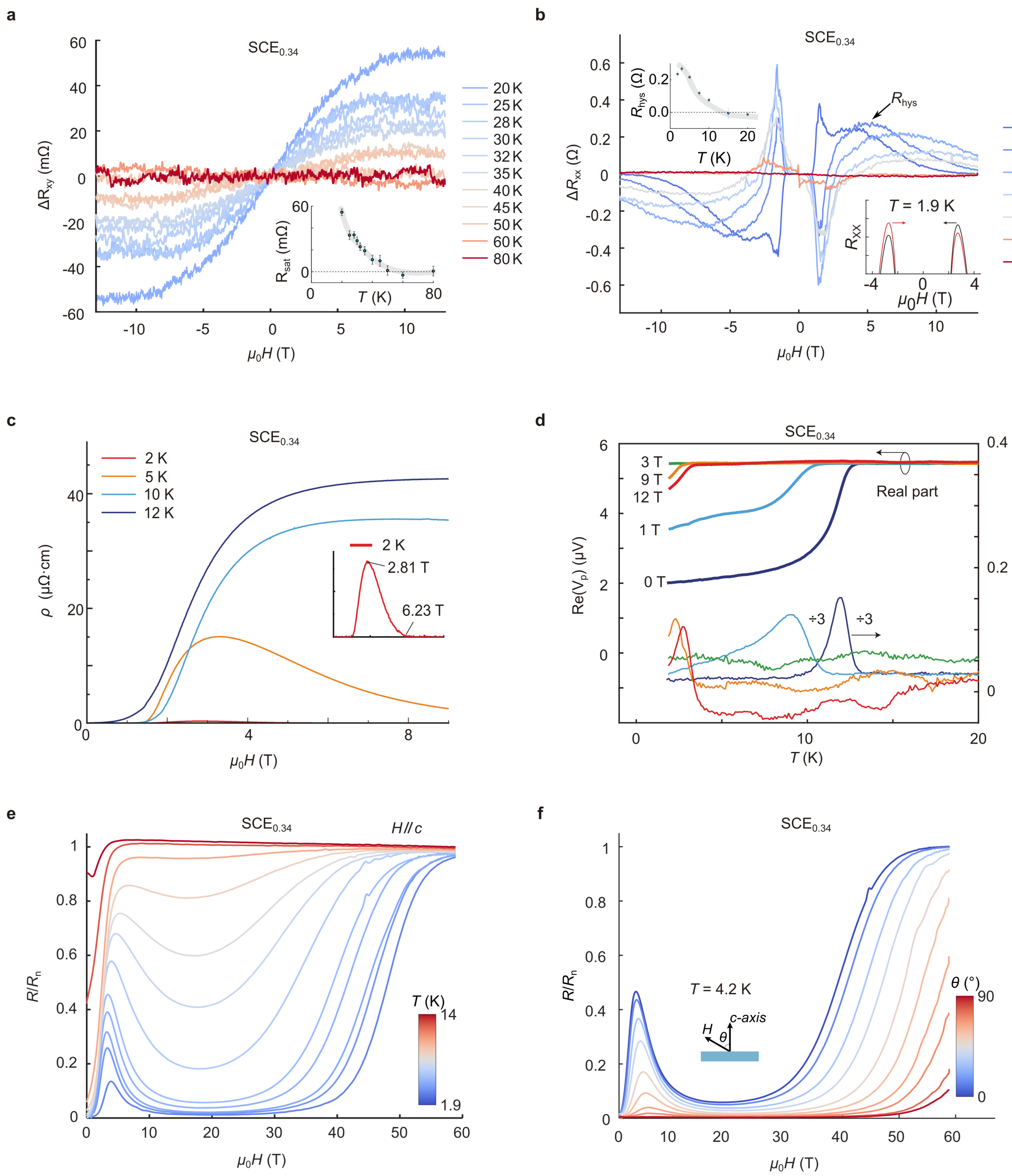

a
SCE0.34
ΔRxy (mΩ)
μ0H (T)
20 K
25 K
28 K
30 K
32 K
35 K
40 K
45 K
50 K
60 K
80 K
Rsat (mΩ)
T (K)
b
SCE0.34
ΔRxx (Ω)
Rhys (Ω)
T (K)
Rhys
T = 1.9 K
Rxx
μ0H (T)
1.9 K
3 K
5 K
7.5 K
10 K
15 K
20 K
c
SCE0.34
2 K
5 K
10 K
12 K
ρ (μΩ·cm)
2.81 T
6.23 T
μ0H (T)
d
SCE0.34
3 T
9 T
12 T
1 T
0 T
Real part
Re(Vp) (μV)
Im(Vp) (μV)
÷3
T (K)
e
SCE0.34
H // c
R/Rn
T (K)
14
1.9
μ0H (T)
f
SCE0.34
T = 4.2 K
H
θ
c-axis
θ (°)
90
0
R/Rn
μ0H (T)

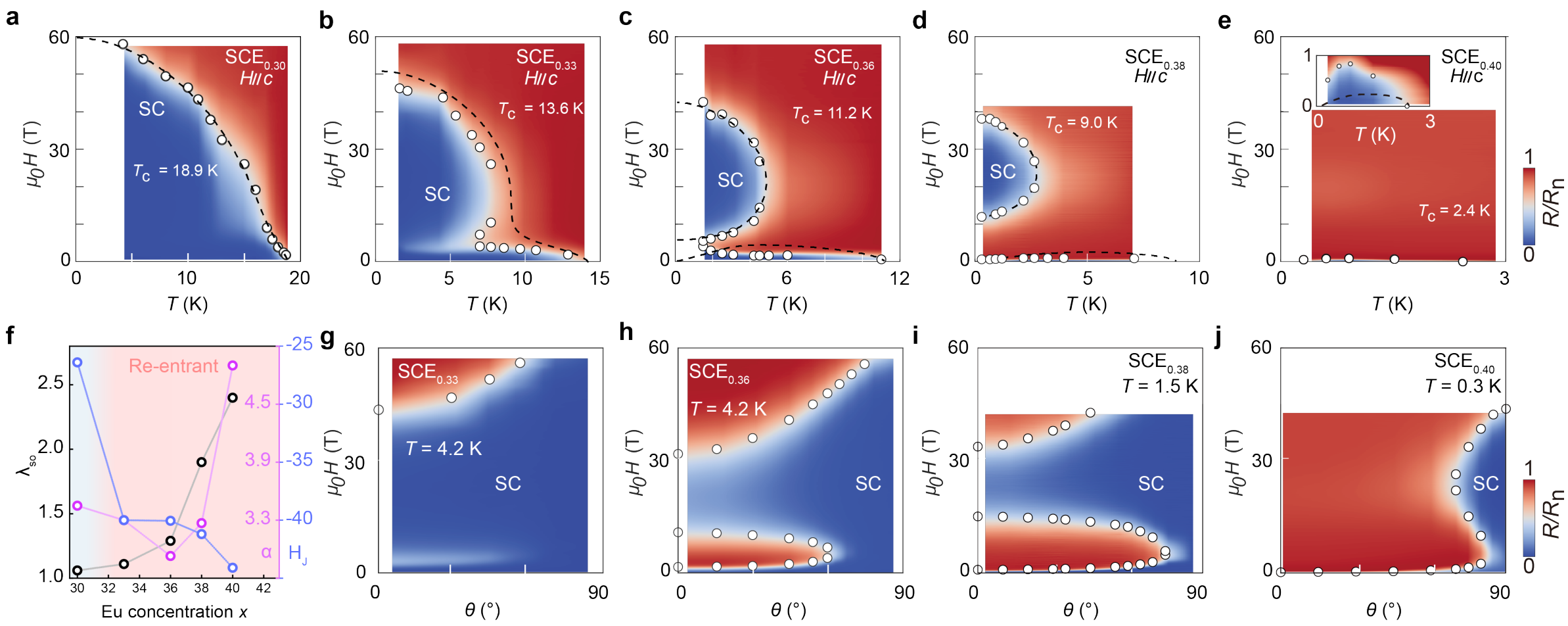

a
b
c
d
e
f
g
h
i
j
SCE0.30
SCE0.33
SCE0.36
SCE0.38
SCE0.40
H//c
SC
Tc = 18.9 K
Tc = 13.6 K
Tc = 11.2 K
Tc = 9.0 K
Tc = 2.4 K
μ0H (T)
T (K)
R/Rn
Re-entrant
λso
α
HJ
Eu concentration x
T = 4.2 K
T = 1.5 K
T = 0.3 K
θ (°)

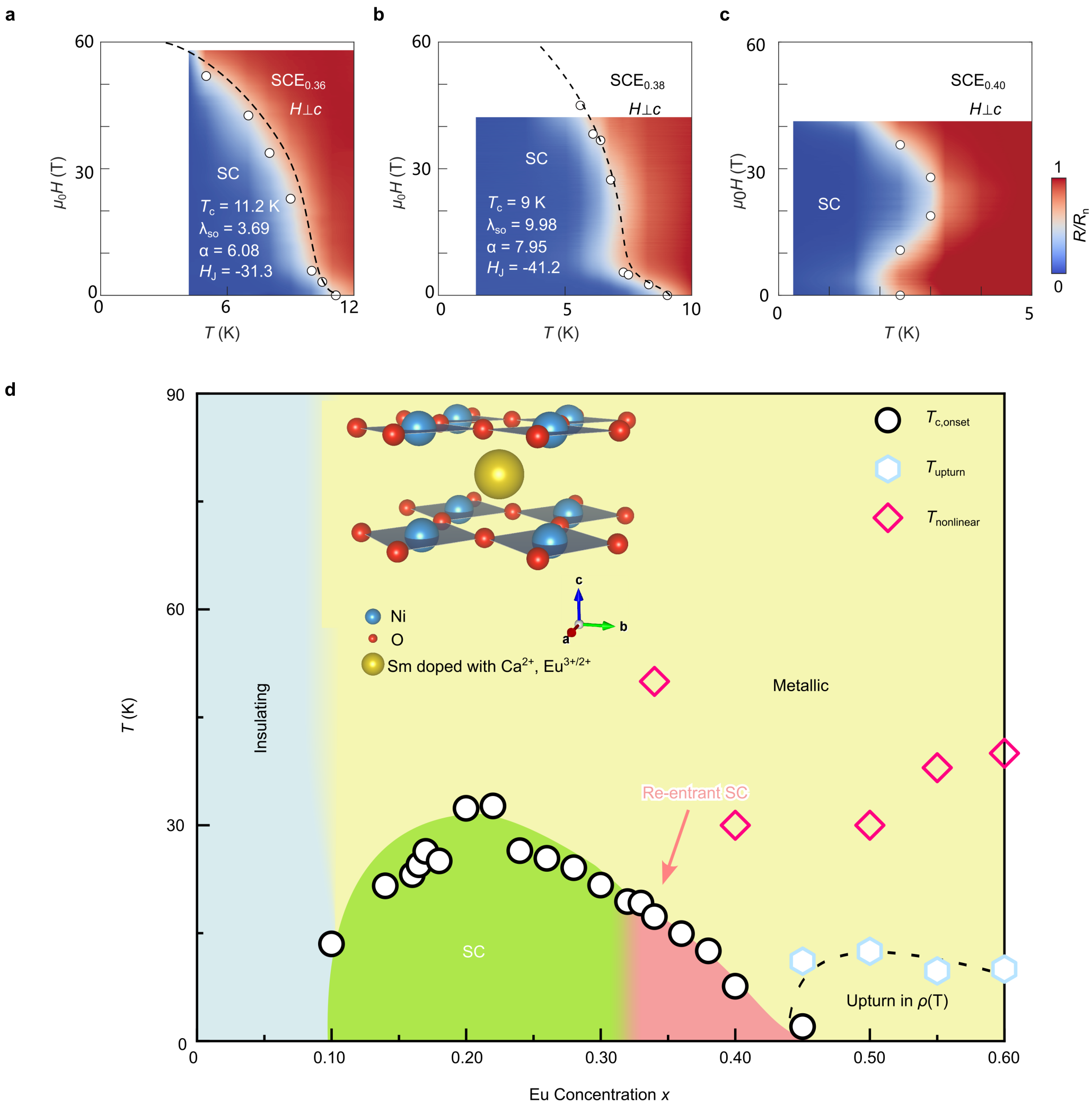
a
SCE0.36
H⊥c
SC
Tc = 11.2 K
λso = 3.69
α = 6.08
HJ = -31.3
μ0H (T)
T (K)
b
SCE0.38
H⊥c
SC
Tc = 9 K
λso = 9.98
α = 7.95
HJ = -41.2
c
SCE0.40
H⊥c
SC
R/Rn
d
Ni
O
Sm doped with Ca2+, Eu3+/2+
Tc,onset
Tupturn
Tnonlinear
Insulating
Metallic
Re-entrant SC
SC
Upturn in ρ(T)
T (K)
Eu Concentration x

## Extended Data Figures

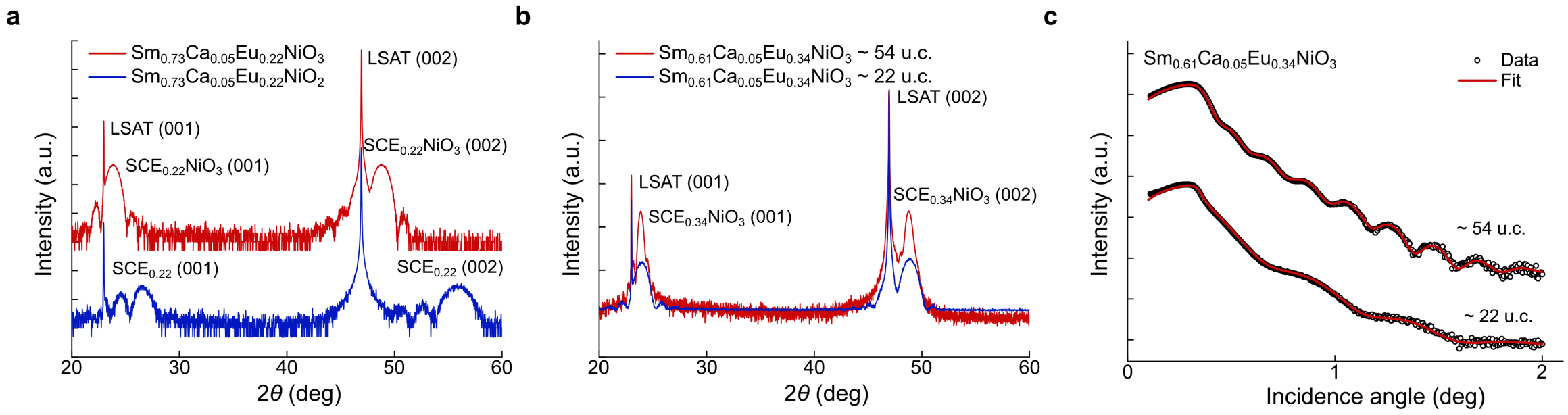

**Extended Data Fig. 1 | The structural characterisations of the samples. a** and **b**, X-ray diffraction (XRD) $\theta$–$2\theta$ symmetric patterns of $Sm_{0.73}Ca_{0.05}Eu_{0.22}NiO_{3(2)}$ and $Sm_{0.61}Ca_{0.05}Eu_{0.34}NiO_3$ thin films. The presence of prominent (00l) film peaks with finite-size oscillations and their peak positions all suggest good film crystallinity across wide ranges of doping and film thickness. c, X-ray reflectivity (XRR) patterns for $Sm_{0.61}Ca_{0.05}Eu_{0.34}NiO_3$ thin films with thicknesses of ~ 22 u.c. and ~ 54 u.c. In experiments, the infinite-layer phase samples with varying doping levels all have a thickness of ~ 22 u.c. The sample $SCE_{0.34}$ used for the mutual inductance two-coil measurements has a thickness of ~ 54 u.c. The XRR fits in Extended Data Fig. 1 were obtained using a standard Parratt formalism as implemented in the commercial analysis package (Rigaku GlobalFit). The ~ 5 u.c. $SrTiO_3$ capping layer was included in the XRR model as the topmost layer. Its thickness and roughness were refined together with those of the infinite-layer film.

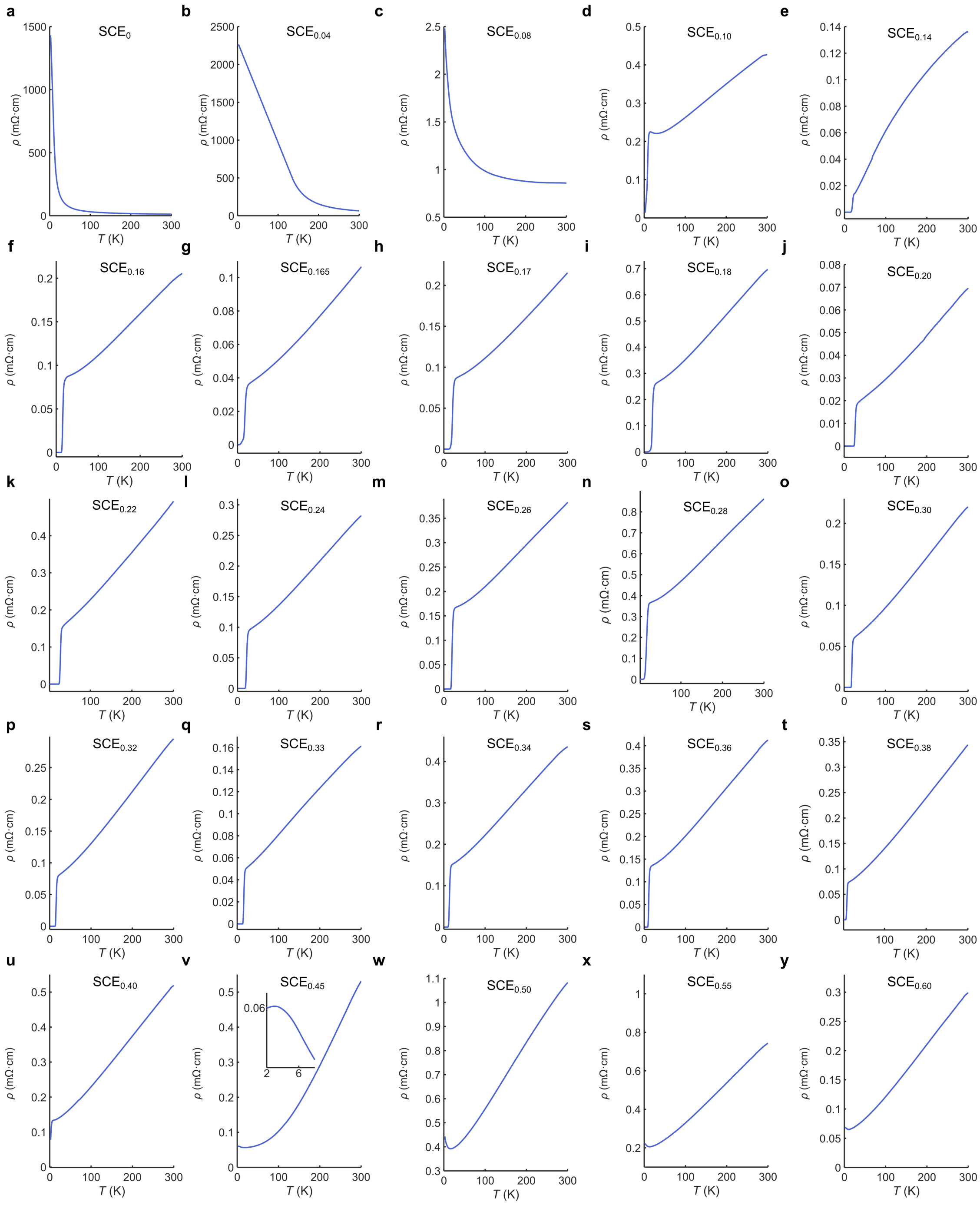

**Extended Data Fig. 2 | Individual $\rho(T)$ curves of $Sm_{0.95\text{-}x}Ca_{0.05}Eu_xNiO_2$ (SCE$_x$) across all doping levels.** The inset of Panel **v** is the zoom-in curve at low temperatures, showing a small downturn in resistivity near $T = 2.3$ K.

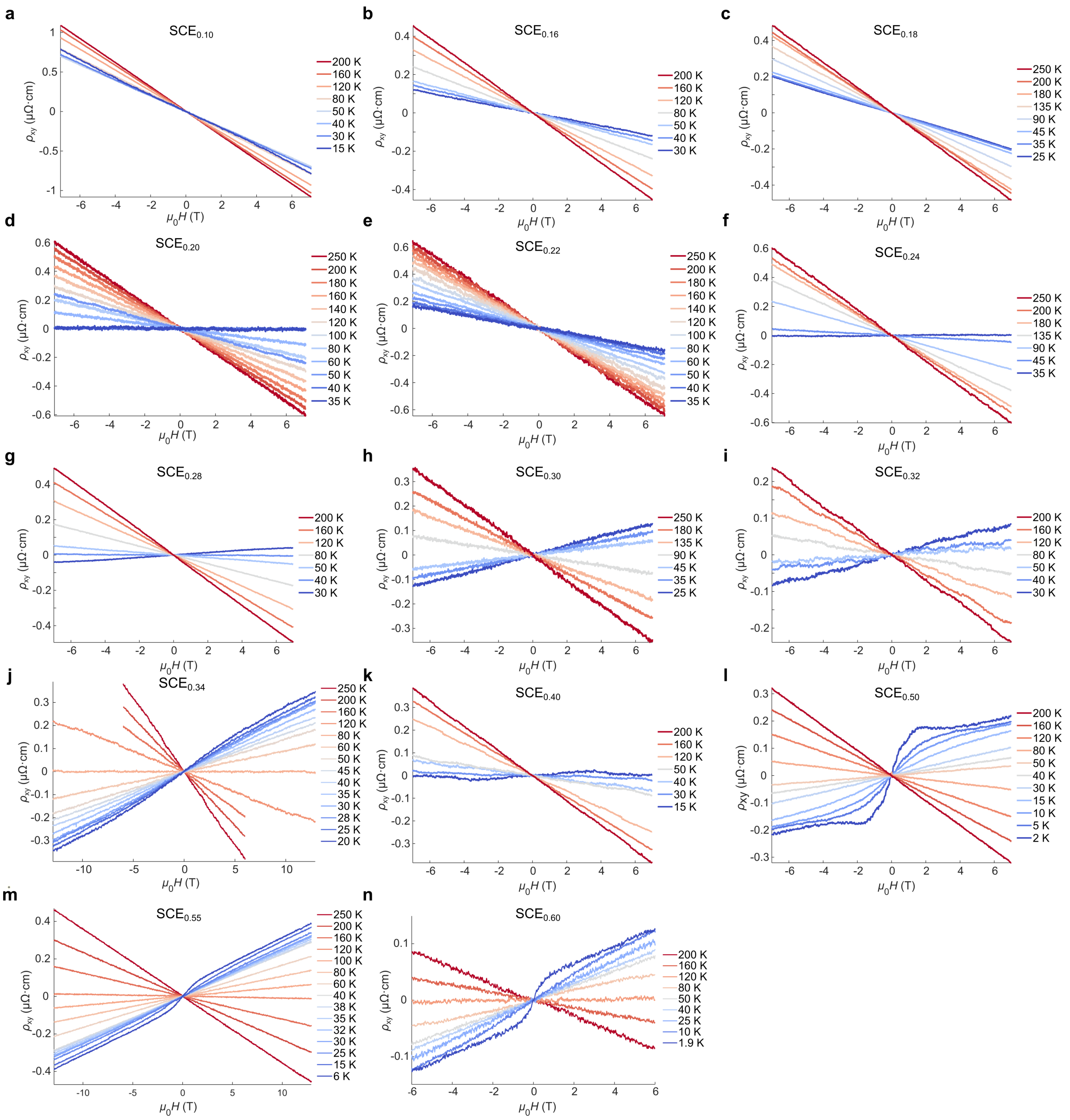

**Extended Data Fig. 3 | Hall resistivity of SCE$_x$ samples across different regimes.** No sign change in Hall coefficient ($R_\mathrm{H}$, slope of $\rho_{yx}(H)$) for $x$ lower than 0.24; Sign change in $R_\mathrm{H}$ for $x$ over 0.24; Nonlinear Hall resistivity at low temperature for $x$ over 0.34.

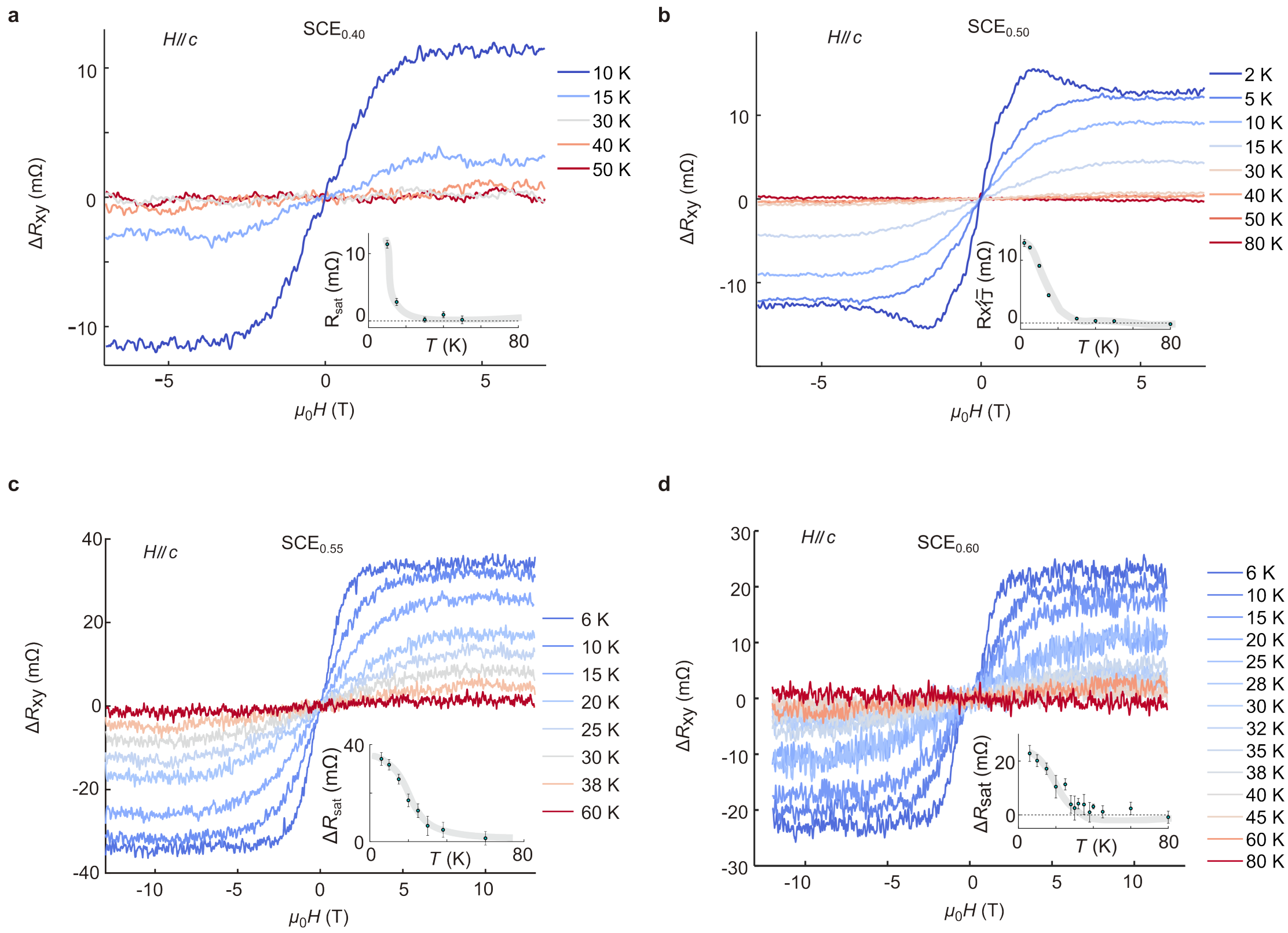

**Extended Data Fig. 4 | Non-linear Hall resistance of SCE$_x$ samples, $x$ = 0.40, 0.50, 0.55 and 0.60.** The insets show the temperature dependence of the saturation resistance, with the grey curves serving as guides to the eye.

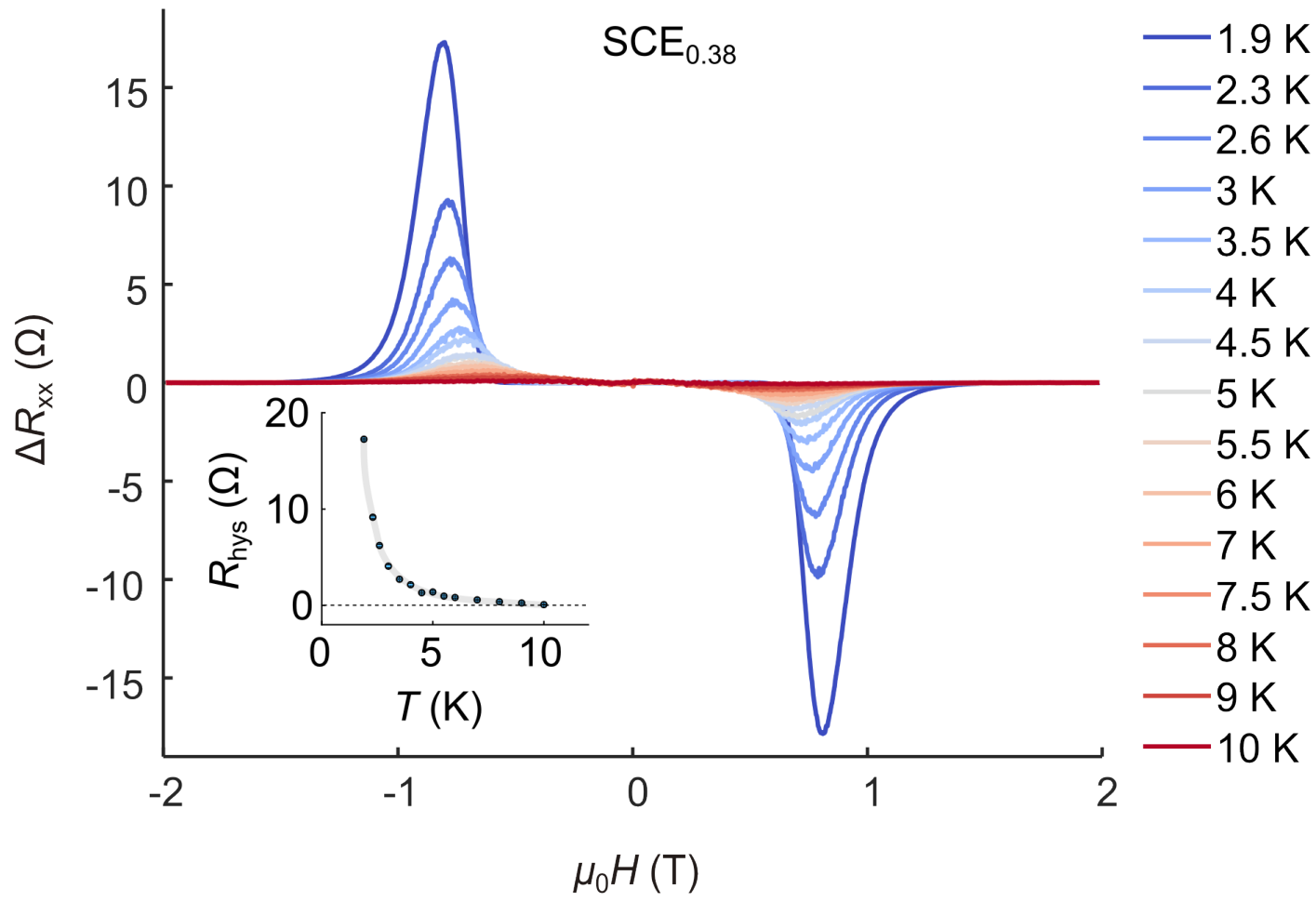

**Extended Data Fig. 5 | Magnetoresistance hysteresis of a $SCE_{0.38}$ sample at different temperatures.** The insets show the temperature dependence of the peak value $R_{hys}$ of the hysteresis curve, with the grey curves serving as a guide to the eye.

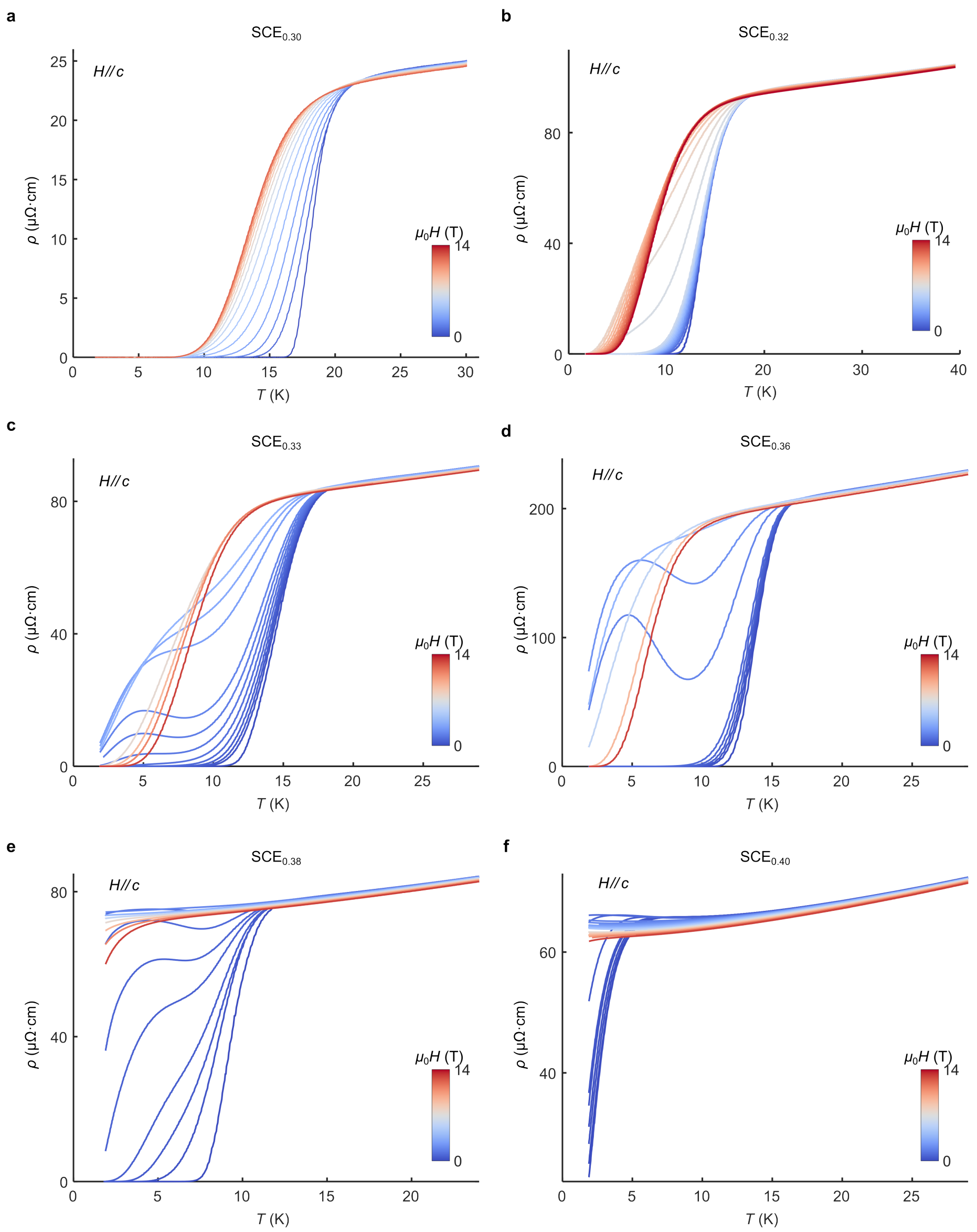

**Extended Data Fig. 6 | Magnetic-field responses of SCE$_x$ samples across different regimes.** Clear re-entrant behaviour starts to show up as $x$ exceeds 0.32.

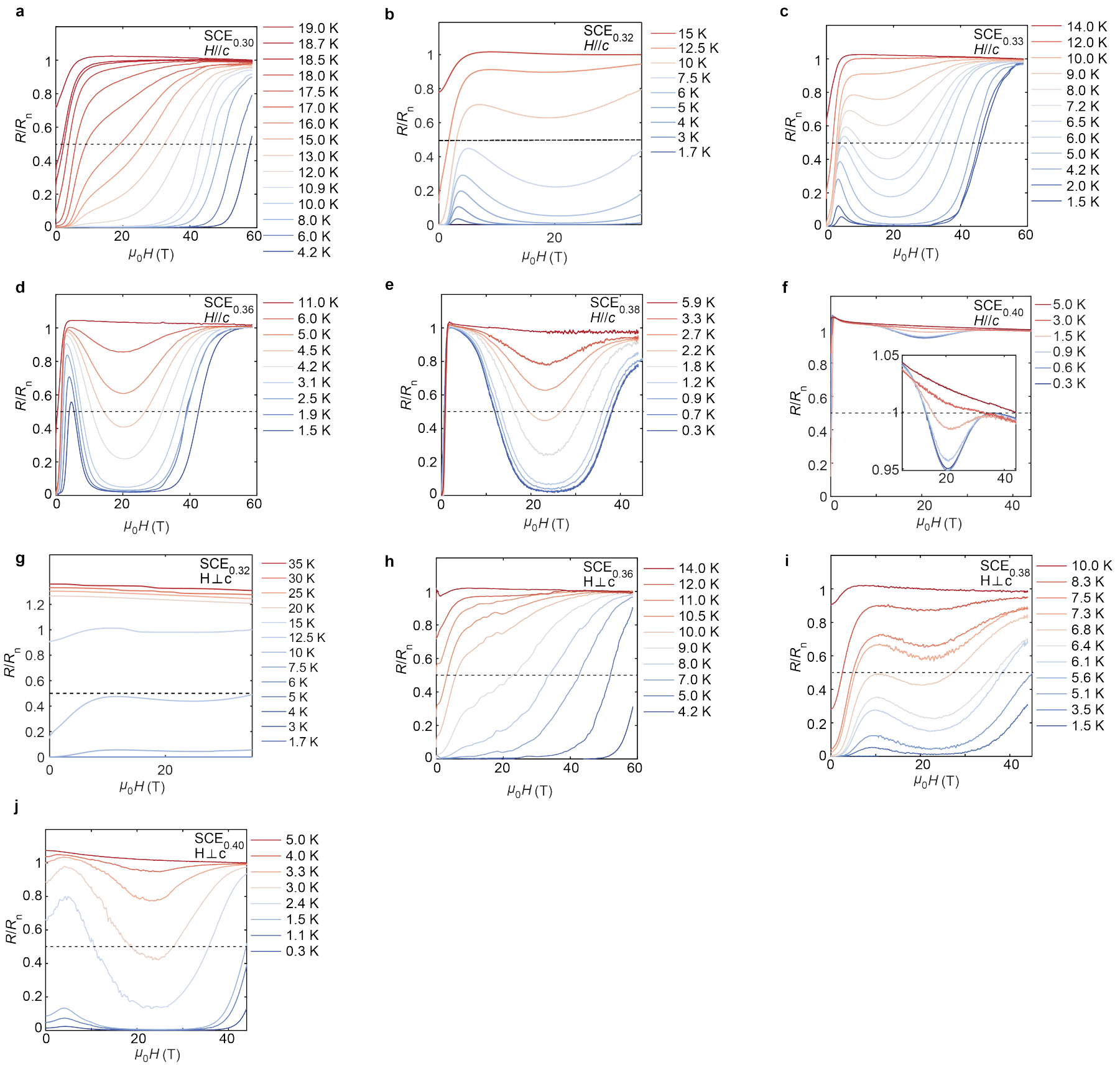

**Extended Data Fig. 7 | High field $\rho(H)$ at different temperatures for SCE$_x$ samples under out-of-plane and in-plane fields.** Here, the $\rho(H)$ curves for SCE$_{0.30}$ show no re-entrant behaviour up to 60 T at $H//c$, clearly demarcating a lower limit of the re-entrant regime. Typical re-entrant behaviour is clearly exhibited for the composition range of $0.32 \leq x \leq 0.38$ at $H//c$. The $\rho(H)$ curves for SCE$_{0.36}$ show no re-entrant behaviour up to 60 T at $H\perp c$. The $\rho(H)$ curves of SCE$_{0.38}$ and SCE$_{0.40}$ under in-plane field show a sign of field enhanced superconductivity.

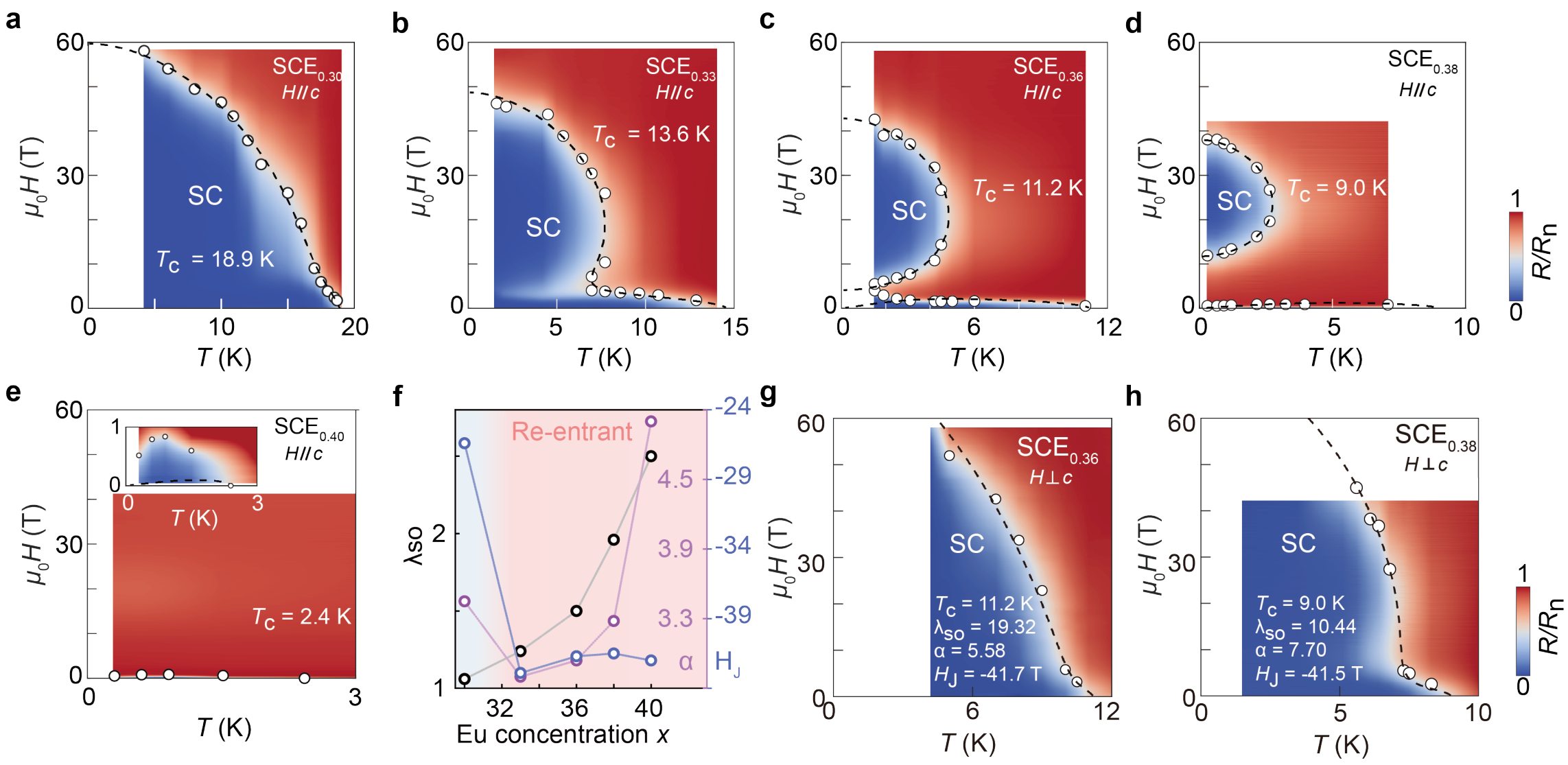

**Extended Data Fig. 8 | Empirically modified fits to the high-field superconducting phase diagrams. a-e**, *H-T-R* contour maps for $SCE_x$ with $x$ = 0.30, 0.33, 0.36, 0.38 and 0.40 for *H*//*c*, reproduced from the main analysis for direct comparison. **f**, Composition dependence of the corresponding fit parameters. The empirical modification improves the agreement for intermediate compositions ($x$ = 0.33 - 0.38). **g,h**, *H-T-R* contour maps for $SCE_x$ with $x$ = 0.36 and 0.38 for *H*⊥*c*. All black dashed lines show fits obtained after introducing an empirical factor of 2 in the argument of the Brillouin function.

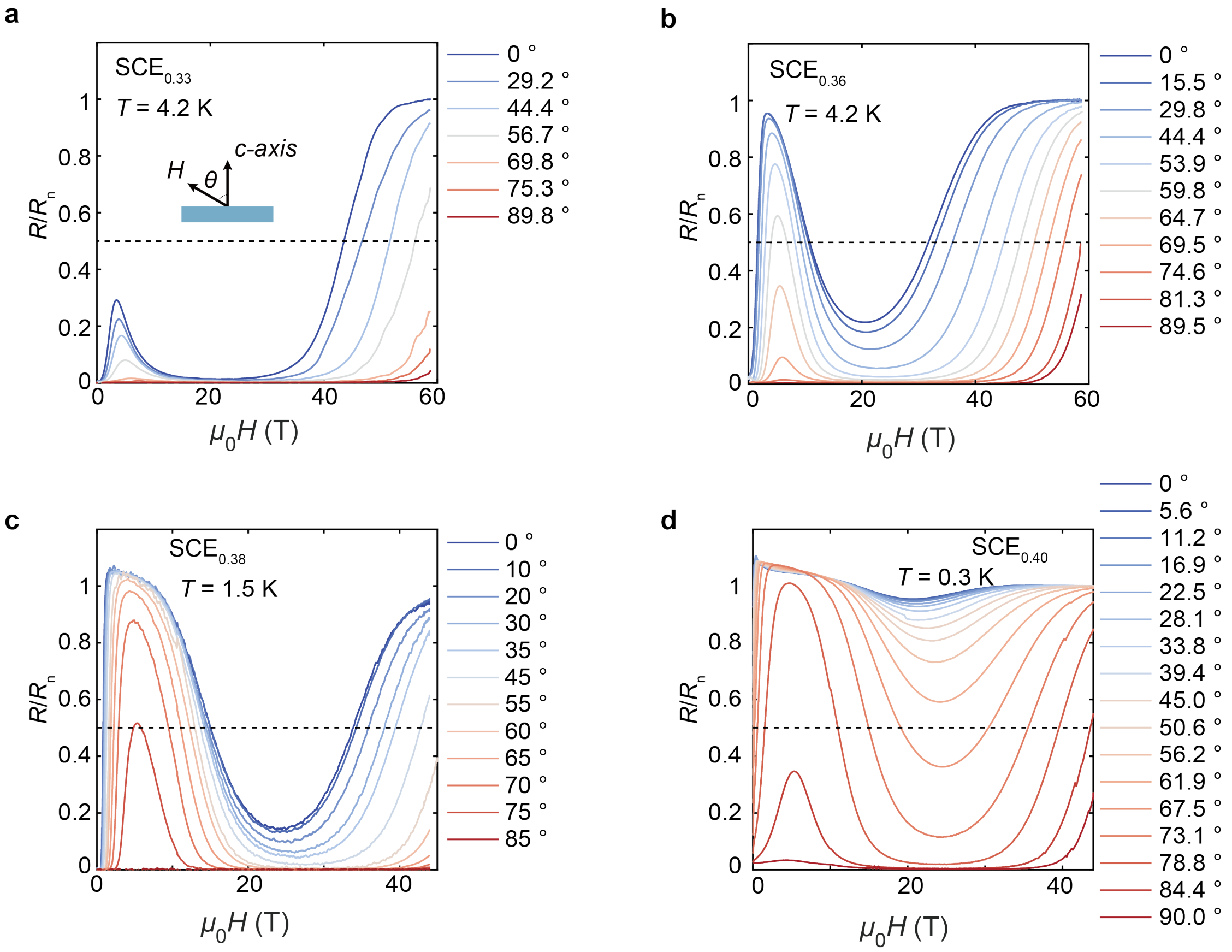

**Extended Data Fig. 9 | $\rho(H)$ up to 45 T with different $\theta$ angles for SCE$_x$ samples.** $\theta$ is the angle between $H$ and the sample surface normal.

| Samples | $T_c$ | $H_P$ | $H_{c2}^{orb}$ | $H_J$ | $\lambda_{so}$ | $\alpha$ |
|---|---|---|---|---|---|---|
| $x = 0.30$ | 18.9 K | 34.8 T | 84.9 T<br>(86.2 T) | -26.4<br>(-26.4 T) | 1.06<br>(1.06) | 3.45<br>(3.45) |
| $x = 0.33$ | 13.6 K | 25.0 T | 51.8 T<br>(49.5 T) | -39.9<br>(-42.9 T) | 1.11<br>(1.24) | 3.3<br>(2.90) |
| $x = 0.36$ | 11.2 K | 20.6 T | 42.9 T<br>(42.9 T) | -40.05<br>(-41.7 T) | 1.29<br>(1.50) | 2.93<br>(2.94) |
| $x = 0.38$ | 9.0 K | 16.6 T | 38.3 T<br>(38.4 T) | -41.2<br>(-41.5 T) | 1.9<br>(1.96) | 3.27<br>(3.28) |
| $x = 0.40$ | 2.4 K | 4.4 T | 15.6 T<br>(15.6 T) | -44.1<br>(-42.1 T) | 2.4<br>(2.5) | 4.9<br>(5.01) |

**Extended Data Table 1 | $H_{c2}^{orb}$ and phenomenological fit parameters for temperature-dependent $H_{c2}$ of different samples.** $H_J$, $\lambda_{so}$, and $\alpha$ are obtained from fits to the unmodified Fisher-extension WHH model. Values in parentheses were obtained using the empirical modified Brillouin form with an extra factor of 2 in the argument.